\documentclass[aps,prl,amsmath,floats,floatfix, twocolumn,
superscriptaddress,nofootinbib,showpacs]{revtex4-1}

\usepackage[colorlinks, pdfborder={0 0 0}, plainpages=false]{hyperref}
\usepackage{graphicx}
\usepackage{xspace}
\usepackage[usenames,dvipsnames]{color}
\usepackage{amssymb}
\usepackage{mathrsfs}
\usepackage{multirow}
\usepackage{comment}
\usepackage[normalem]{ulem}

\usepackage{ulem}
\newcommand{\Caltech}{\affiliation{TAPIR,
    Walter Burke Institute for Theoretical Physics, 
    California Institute of Technology, Pasadena, CA 91125, USA}}
\newcommand{\Cornell}{\affiliation{Center for Radiophysics and Space
    Research, Cornell University, Ithaca, New York 14853, USA}}
\newcommand{\UCSD}{\affiliation{Center for Astrophysics and Space Sciences, 
                   Center for Computational Mathematics, 
                   San Diego Supercomputer Center, 
                   University of California San Diego, 
                   9500 Gilman Drive, La Jolla, California 92093-0424, USA}}

\begin{document}

\title{Fast and accurate prediction of numerical relativity waveforms from 
binary black hole coalescences using surrogate models}

\author{Jonathan Blackman} \Caltech
\author{Scott E.~Field} \Cornell
\author{Chad R.~Galley} \Caltech
\author{B\'{e}la Szil\'{a}gyi} \Caltech 
\author{Mark A.~Scheel} \Caltech 
\author{Manuel Tiglio} \UCSD
\author{Daniel A.~Hemberger} \Caltech

\date{\today}

\begin{abstract}
Simulating a binary black hole (BBH) coalescence by solving Einstein's equations 
is computationally expensive, requiring days to months of  
supercomputing time. 
Using reduced order modeling techniques, 
we construct 
an accurate
surrogate model, 
which is evaluated in a millisecond to a second,
for numerical relativity (NR) waveforms
from non-spinning BBH coalescences with 
mass ratios in $[1, 10]$ 
and durations corresponding to about
$15$ orbits
before merger. 
We assess the model's uncertainty and show that our modeling strategy
 predicts NR waveforms {\em not} used for the surrogate's training
 with errors nearly as small as the numerical error of the NR code.
Our model includes all spherical-harmonic 
${}_{-2}Y_{\ell m}$
waveform modes resolved by the NR code up to $\ell=8.$
We compare our surrogate model to 
Effective One Body waveforms from $50$-$300 M_\odot$ for advanced LIGO detectors
and find that the surrogate is always more faithful (by at least an order of magnitude in most cases).
\end{abstract}


\maketitle

Since the breakthroughs of 
2005~\cite{Pretorius2005a, Campanelli2006a, Baker2006a}, tremendous progress in 
numerical relativity (NR) has led to 
hundreds of simulations of  binary black hole (BBH) 
coalescences~\cite{Aylott:2009ya, Ajith:2012az, Mroue:2013PRL, 
Hinder:2013oqa, Pekowsky:2013ska, Aasi:2014tra, SXSCatalog}.
This progress has been driven partly by
data analysis needs of advanced ground-based gravitational wave detectors like 
 LIGO \cite{Harry2010} and  Virgo \cite{aVirgo}. 
Recent upgrades to these detectors 
are expected to yield the first direct 
detections of gravitational waves (GWs) 
from compact binary coalescences~\cite{Abadie:2010cf}.

Despite the remarkable progress of the NR community, a single high-quality 
simulation typically requires days to months of supercomputing time.
This high computational cost makes it difficult to directly use NR waveforms for data analysis, 
except for injection studies~\cite{Aylott:2009ya, Aasi:2014tra}, 
since detecting GWs and inferring their source parameters may require thousands to
millions of accurate gravitational waveforms.
Nevertheless, a first template bank for nonspinning
binaries in Advanced LIGO
has been recently constructed from NR waveforms~\cite{Kumar:2013}. Furthermore,
NR waveforms have been used successfully
in calibrating 
inspiral-merger-ringdown (IMR) 
effective-one-body~(EOB)~\cite{Buonanno99, 
Damour2009a, 
Buonanno:2009qa, Pan:2009wj,
Pan:2011gk, 
Damour:2012ky, Taracchini:2012}
and 
phenomenological~\cite{Ajith-Babak-Chen-etal:2007,
Santamaria:2010yb, Sturani:2010yv, Hannam:2013oca}
models.
These models have free parameters that can be set by matching to NR waveforms
and are suitable for 
certain GW data analysis studies~\cite{Abadie:2011kd}.
However, these models can have systematic errors since they 
assume {\em a priori} physical waveform structure and 
are calibrated and tested against a small set of NR simulations.

\begin{figure}
        \includegraphics[width=\columnwidth, trim=30 20 22 0]{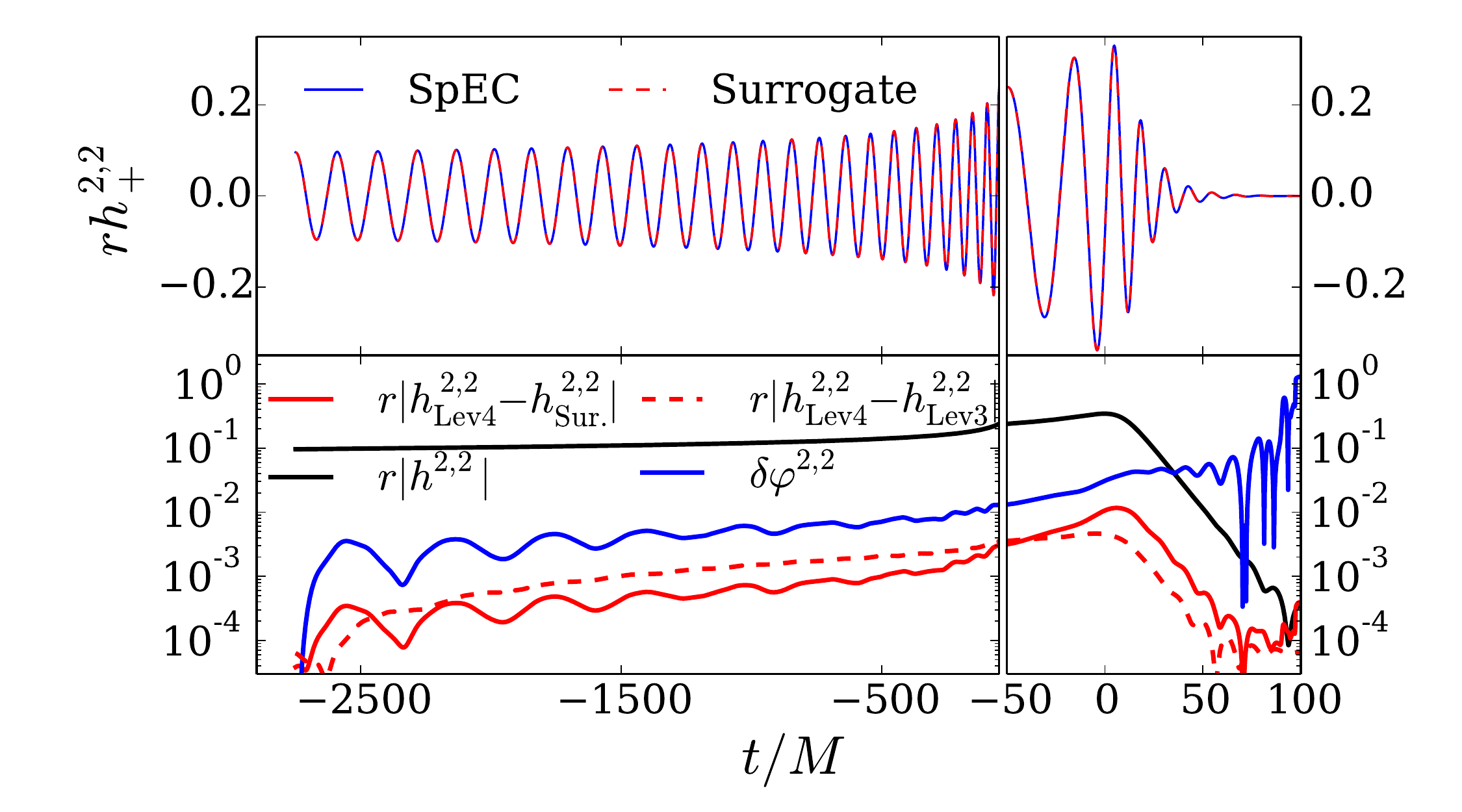}
        \caption{
                {\bf Top}: The $+$ polarization $(2,2)$ mode prediction for 
                $q=2$, the surrogate model's worst 
                prediction over $q$ from a ``leave-one-out''
                surrogate that was {\em not} trained with this waveform (see below).
                Our full surrogate, trained on the entire data set,
                is more accurate.
                {\bf Bottom}: Phase $\delta\varphi^{2,2}$ and waveform differences 
                between the surrogate and 
                highest resolution (Lev4) SpEC waveforms. Also shown is 
                the SpEC numerical truncation error found by comparing the 
                two highest resolution (Lev4 and Lev3) waveforms.
        }
        \label{fig:waveform}
\end{figure}

In this Letter, we present an {\it ab initio} methodology based on 
surrogate~\cite{Field:2013cfa, brown_sc_2013_13} and reduced order 
modeling techniques~\cite{Maday:2002,Veroy:2003,prudhomme:2002,Chen:2010,Quarteroni:2011} that 
is capable of accurately predicting the gravitational waveform 
outputs from NR {\it without} any phenomenological 
assumptions or approximations to general relativity.
From a small set of specially selected non-spinning BBH simulations performed with 
the Spectral Einstein Code 
(SpEC)~\cite{SpECwebsite,Scheel2014,Szilagyi:2014fna},
we build a surrogate 
model that can be used in place of performing SpEC simulations.
The techniques are general, however, and directly apply to other NR codes or even analytical waveform models.
The surrogate model constructed here 
generates non-spinning BBH waveforms with mass ratios $q \in [1,10]$,
contains $25$--$31$ gravitational wave cycles before peak amplitude, and
includes
many spherical-harmonic modes
(see Table~\ref{tab:mode_err} and its caption).
These choices are made based on available NR waveforms and are not limitations of the method.
Our surrogate model has errors close to the estimated numerical error of the input
waveforms.
An example comparing the surrogate output to an NR waveform can be seen in Fig.~\ref{fig:waveform}.
This simulation took $9.3$ days using $48$ cores but only $\sim 0.01$ sec for the surrogate evaluation of the (2,2) mode.

Previous work~\cite{Field:2013cfa, Purrer:2014}
built surrogates for EOB waveforms; 
building and assessing surrogate models of NR waveforms have 
unique challenges associated with
input waveforms that are expensive to compute.
We summarize next the construction of our 
model, focusing on steps 
not addressed in \cite{Field:2013cfa}
but are required for NR surrogates.

{\it Parametric sampling--} 
Typically, a surrogate model is trained on a dense
set of waveforms known as the {\em training set}.
In the case of NR, we cannot afford to generate a large number of waveforms.
Instead, we 
generate a dense
set of non-spinning waveforms using an EOB
model~\cite{Pan:2009wj}, as implemented in ~\cite{LAL}, which 
contains the  $(\ell,m) = \{(2,2), (2,1), (3,3), (4,4),(5,5)\}$ spin-weight $-2$
spherical-harmonic modes and captures robust features of NR waveforms.
The EOB training set waveforms 
are computed for times in $[-2750, 100]M$ ($M$ is the total mass), 
which is the interval over which we build our surrogates.

Next, on this training set
we apply a
greedy algorithm to expose the most relevant
mass ratio values ~\cite{Binev10convergencerates, Field:2011mf}.
The algorithm proceeds from a linear basis constructed from $i$ waveforms already chosen.
The $L^2$ norms of the differences between the training set waveforms and their projection onto this basis are computed. The waveform with the largest such error is added to the basis as its $i+1$ element.
SpEC simulations of non-spinning BBH mergers are then performed for these mass ratios.
The resulting NR waveforms are used to build our surrogates without any further input from the EOB model.

We seeded the greedy algorithm with $5$ publicly available SpEC simulations 
of non-spinning BBH mergers ~\cite{Pan:2011gk,SXSCatalog}
(see Table~\ref{tab:runs}),
and the next $17$ (ordered) mass ratio values are the algorithm's 
output based on the EOB model. The final 
$\sim$$10$ mass ratios 
are included to improve 
the surrogate if necessary, 
since we can 
assess the surrogate model's accuracy only after it is built. 
Our method for building surrogates is hierarchical~\cite{Field:2011mf, Field:2013cfa}; 
additional NR waveforms can be included to improve the model's accuracy.

{\it Generating the NR waveforms--} 
Table \ref{tab:runs} summarizes the $22$ SpEC simulations
used in this paper. See, e.g., Ref.~\cite{Scheel2014} for the 
numerical techniques used in SpEC. 
The numerical resolution is denoted by ``Lev$i$'', where $i$ is
an integer that controls the local truncation error 
in the metric and its derivatives allowed by adaptive
mesh refinement 
(AMR)
in SpEC; larger numbers correspond to smaller
errors (the error threshold scales like $e^{-i}$) 
and more computationally-expensive simulations.
The scaling of
global quantities (e.g. waveform errors) with $i$ is difficult to estimate
{\it a priori}.
Two to five levels of resolution
are simulated for each mass ratio.
To achieve quasi-circular orbits, initial data are subject to 
an iterative eccentricity reduction procedure resulting 
in eccentricities $\lesssim 7\times 10^{-4}$ \cite{Pfeiffer-Brown-etal:2007, Buonanno:2010yk, Mroue:2012kv}.

\begin{table}
\begin{tabular}{c | c c c c c|| c | c c c c c}
$\#$ & ID & $q$ & $e_{-5}$ & $T/M$ & Orbs & $\#$ & ID & $q$ & $e_{-5}$ & $T/M$ & Orbs \\
\hline
$1$ & $180$ & $1.00$ & $5.1$ & $9867$ & $28.2$ & $12$ & $191$  & $2.51$ & $65$  & $6645$ & $22.5$\\
$2$ & $181$ & $6.00$ & $5.8$ & $7056$ & $26.5$ & $13$ & $192$ & $6.58$ & $4.0$ & $5149$ & $21.1$\\
$3$ & $182$ & $4.00$ & $12$  & $3840$ & $15.6$ & $14$ & $193$   & $3.50$ & $3.0$ & $5242$ & $19.6$\\
$4$ & $183$ & $3.00$ & $4.8$ & $4008$ & $15.6$ & $15$ & $194$  & $1.52$ & $74$  & $5774$ & $19.6$\\
$5$ & $184$ & $2.00$ & $15$  & $4201$ & $15.6$ & $16$ & $195$ & $7.76$ & $22$  & $5226$ & $21.9$\\
$6$ & $185$ & $9.99$ & $31$  & $5817$ & $24.9$ & $17$ & $196$ & $9.66$ & $23$  & $5330$ & $23.1$\\
$7$ & $186$ & $8.27$ & $16$  & $5687$ & $23.7$ & $18$ & $197$   & $5.52$ & $25$  & $5061$ & $20.3$\\
$8$ & $187$ & $5.04$ & $3.0$ & $4807$ & $19.2$ & $19$ & $198$  & $1.20$ & $17$  & $6315$ & $20.7$\\
$9$ & $188$ & $7.19$ & $15$  & $5439$ & $22.3$ & $20$ & $199$ & $8.73$ & $8.5$ & $5302$ & $22.6$\\
$10$& $189$ & $9.17$ & $13$  & $6019$ & $25.2$ & $21$ & $200$ & $3.27$ & $36$  & $5507$ & $20.2$\\
$11$& $190$ & $4.50$ & $2.5$ & $5199$ & $20.1$ & $22$ & $201$ & $2.32$ & $15$  & $5719$ & $20.0$
\end{tabular}
\caption{Properties of the highest resolution SpEC simulations used for building BBH waveform surrogates. 
         The quantity $e_{-5}$ is the orbital eccentricity divided by $10^5$ \cite{Mroue:2012kv}.
         The duration $T/M$ and number of orbits (Orbs) are also given.  
         The SpEC simulations are available in the public waveform 
         catalog~\cite{SXSCatalog} under the name ``SXS:BBH:{\it ID}.''
	}
   \label{tab:runs}
\end{table}

SpEC numerically solves an initial boundary value problem defined on a finite 
computational domain. 
To obtain waveforms at future null infinity $\mathscr{I}^+$, we use the 
Cauchy characteristic extraction (CCE) 
method~\cite{Bishop:1997ik, Babiuc:2010ze, Winicour2009, Reisswig:2010di, Taylor:2013zia}.
Using the PittNull code \cite{Bishop:1997ik, Babiuc:2010ze, Winicour2009}, 
we compute the Newman-Penrose scalar $\Psi_4$ at $\mathscr{I}^+$
and finally obtain
the gravitational wave strain $h$
through two temporal integrations. We minimize the low-frequency, noise-induced ``drifts"~\cite{Reisswig:2010di}
by using frequency cut-offs.\footnote{We integrate $\Psi_4$ twice in the (dimensionless) frequency domain
by dividing $-\Psi_{4}^{\ell,m}(f)$ by $[ 2\pi\max(f, 2 f_0 /3) ]^2$, 
where $f_0$ is the initial GW mode frequency.}

\begin{figure}
        \includegraphics[width=\columnwidth, trim=15 30 20 18]{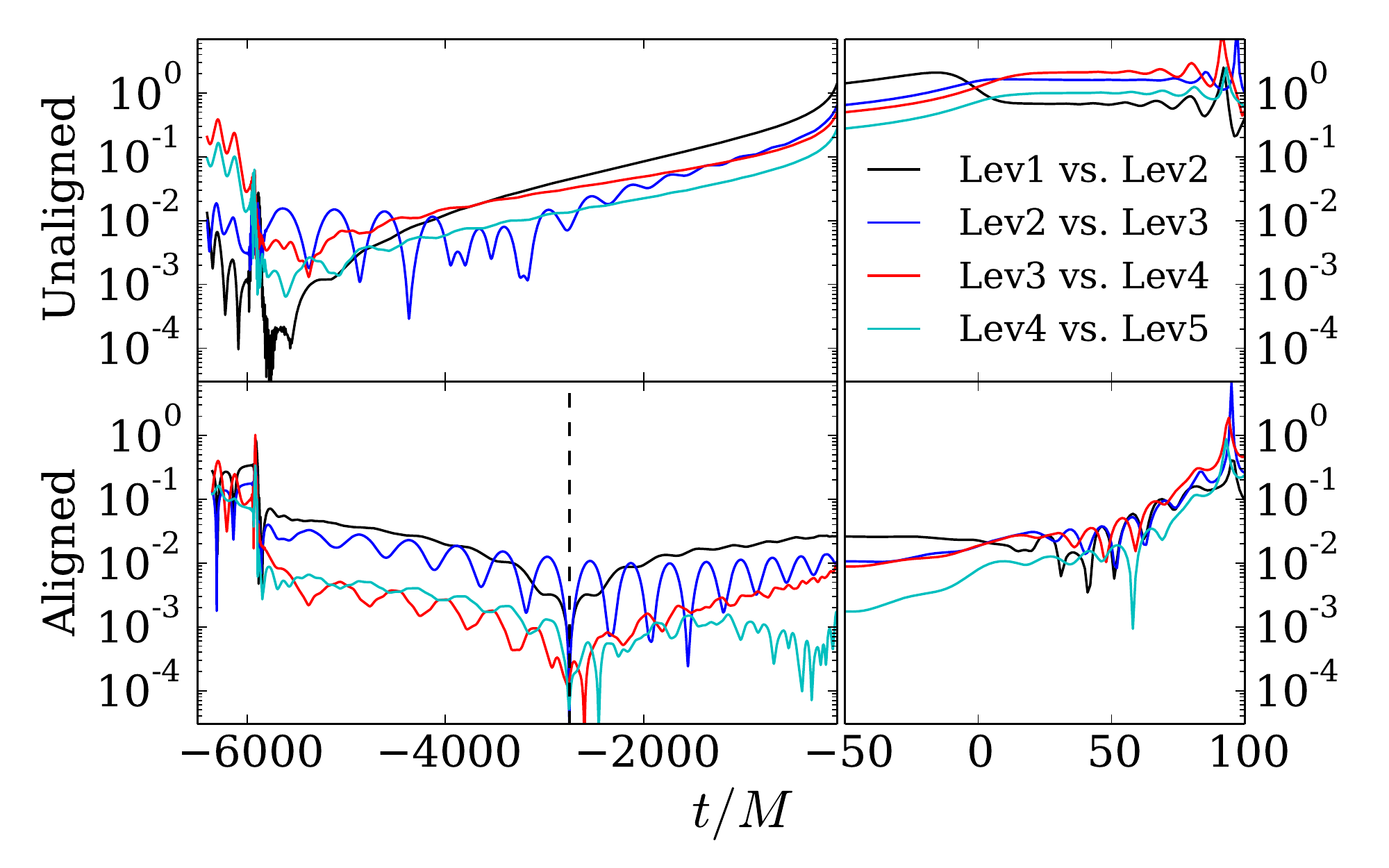}
        \caption{
        The relative error, $|h^{22}_i - h^{22}_{i+1} |/| h^{22}_{i+1}|$, of successive resolutions SpEC Lev$i$
        for the (2,2) mode of simulation $19$ in Table~\ref{tab:runs}.
        {\bf Top}: Waveform output as directly given by SpEC (``Unaligned").
        {\bf Bottom}: 
        ``Aligned,'' which involves a multi-mode peak alignment scheme described by Eq.~\eqref{eq:peakalign}
        followed by a rotation of the binary around the $z$-axis to 
        align the waveform phases at $t_i = -2750M$. Our surrogate is built from 
        NR waveform data after alignment, and so this measurement of truncation error is
        the most relevant for surrogate model building.
        }
        \label{fig:Convergence19}
\end{figure}

\begin{figure}
        \includegraphics[width=\columnwidth, trim=45 20 20 0]{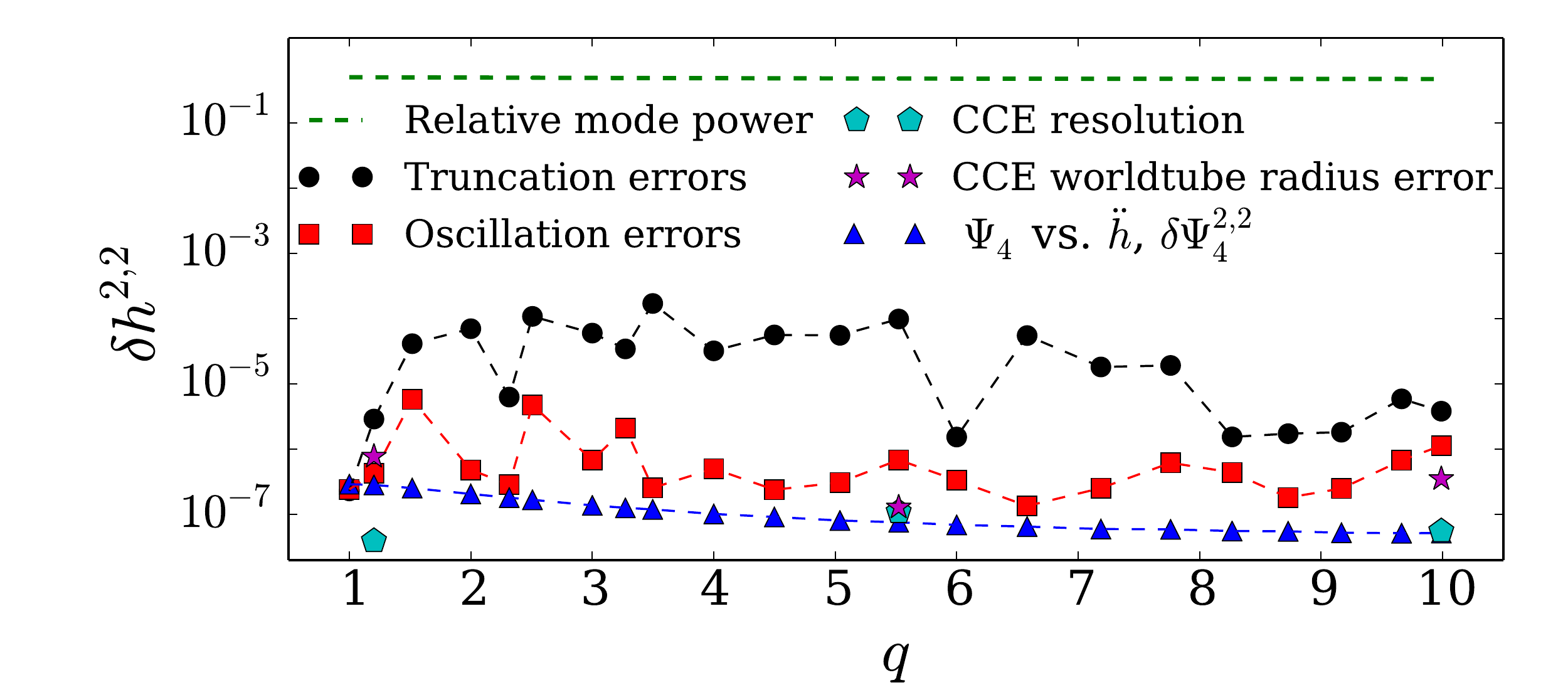}
        \caption{
        Numerical truncation errors (black) 
        dominate 
        all other sources of error for the (2,2) mode, 
        except for simulation $1$ ($q=1$), where the truncation errors are already very small. 
        For some weaker modes, systematic amplitude oscillations primarily due to eccentricity may become more relevant.
        }
        \label{fig:OscillationError}
\end{figure}

Figure~\ref{fig:Convergence19} shows the 
convergence typically observed in our simulations when using AMR.
Because AMR makes independent
decisions for different Lev$i$, a particular subdomain
may sometimes have the same number of grid points for two different
values of Lev$i$ at a given time, and the subdomain boundaries
do not necessarily coincide for different Lev$i$.  Thus, plots 
like Figure~\ref{fig:Convergence19} sometimes show anomalously small
differences between particular pairs of numerical resolutions (for instance
Lev2 vs. Lev3 near $t=-3500M$ 
in the top panel of Figure~\ref{fig:Convergence19}). 
See Sec.~IIIB of \cite{Scheel2014}.
Nevertheless, the waveform differences generally decrease quickly 
with increasing resolution.
Let
\begin{equation}
	\delta h^{\ell,m} (q) \equiv 
\frac{ \| h_1^{\ell,m}(\cdot;q) - h_2^{\ell,m}(\cdot;q) \|^2}{\sum_{\ell,m} \| h_2^{\ell,m}(\cdot;q) \|^2} 
\label{eq:deltahlm}
\end{equation}
be the disagreement between two waveform modes 
$h_1^{\ell,m}$ and $h_2^{\ell,m}$ where $\| h^{\ell,m}(\cdot;q) \|^2 = \int dt \, |h^{\ell,m}(t;q)|^2$.  
We estimate the numerical truncation error of each mode when
$h_1$ and $h_2$ are waveforms computed at the two highest resolutions. 
The full waveform\footnote{
Throughout, we exclude $m=0$ modes because (non-oscillatory) 
Christodoulou memory is not accumulated sufficiently in current NR simulations~\cite{Favata:2008yd}.}
error for a given mass ratio is
$\delta h (q) = \sum_{\ell,m} \delta h^{\ell,m}(q)$. We 
report numerical truncation 
errors after an overall simulation-dependent 
time shift and rotation
(which we shall refer to as 
{\em surrogate alignment}, 
described in the next section), which are physically unimportant coordinate changes.
The resulting estimated numerical truncation errors of the dominant $(2,2)$ modes, using our 
surrogate alignment scheme, 
are shown in Fig.~\ref{fig:OscillationError} (black circles). 

Additional error sources are non-zero eccentricity in the (intended
to be circular) NR initial data,
and an imperfect procedure for integrating $\Psi_4^{\ell, m}$ to 
obtain $h^{\ell, m} \equiv A^{\ell, m} \exp (-i\varphi^{\ell, m} )$.
These both cause
small oscillations in the waveform amplitudes $A^{\ell, m}(t)$
and phases $\varphi^{\ell, m}(t)$~\cite{Reisswig:2010di, Mroue2010}
that we model following~\cite{Mroue2010}.
We also compute the error in the strain integration scheme by comparing $\Psi_{4}^{\ell,m}$ to two time derivatives of $h^{\ell,m}$, 
as well as estimates for numerical errors in the CCE method ~\cite{Taylor:2013zia}.
For the $(2,2)$ mode, these additional errors
are negligibly small compared to 
 SpEC truncation errors (cf.~Fig.~\ref{fig:OscillationError}).

{\it Preparing NR waveforms for surrogate modeling--}
We apply a simulation-dependent time shift and physical rotation about the $z$-axis so that all the 
modes' phases are aligned. This reveals the underlying parametric smoothness 
in $q$ that will be useful for building a surrogate. 
Our time shifts set each waveform's {\it total} amplitude
\begin{equation} \label{eq:peakalign}
A(t; q)^2 \equiv \int_{S^2} \!\!\! d\Omega \, |h(t, \theta,\phi; q)|^2 = \sum_{\ell,m}|h^{\ell,m}(t; q)|^2 \, ,
\end{equation}
to be maximum at $t=0$.
After enforcing this alignment scheme we interpolate the waveform mode 
amplitudes and phases 
onto an array of uniformly spaced times in $[-2750, 100]M$,
with $\Delta t = 0.1M$.
Finally, we align the initial gravitational
wave mode phases by performing a simulation-dependent, 
constant (in time) 
physical rotation 
about the $z$-axis so that
$\varphi^{2,2}(t_i) = \varphi^{2,-2}(t_i)$,
which fixes a physical rotation up to multiples of $\pi$.
We resolve the ambiguity by requiring $\varphi^{2,1}(t_i) \in (-\pi, 0]$. 
Waveform truncation errors, after performing this surrogate alignment scheme, are 
shown in Fig.~\ref{fig:Convergence19}.
In what follows,
we call ``truncation error 
after surrogate alignment" simply ``truncation error."

{\it Building the surrogate--} 
Each 
$m > 0$ mode, $h^{\ell, m}(t;q)$, is modeled separately while 
(due to reflection symmetry about the orbital plane) $m <0$ modes are 
evaluated using $h^{\ell,-m}(t;q) = (-1)^\ell h^{\ell,m}(t;q)^*$.
We model all $m\ne0$ modes 
but keep only those yielding smaller
surrogate errors $\delta h^{\ell,m}$ compared to setting the mode to zero.
Table~\ref{tab:mode_err} lists our modeled modes and their errors.

Our complete surrogate waveform model  is defined by 
$h_{\rm S}(t,\theta,\phi; q) = \sum_{\ell,m} h^{\ell, m}_{\rm S}(t; q) {}_{-2}Y_{\ell m} \left(\theta, \phi \right)$ where
\begin{align}
\begin{split}
	h^{\ell, m}_{\rm S}(t; q) & = A^{\ell, m}_{\rm S} (t; q) e^{-i \varphi^{\ell, m}_{\rm S} (t; q)}  \, , \\
	X_{\rm S}^{\ell, m} (t; q) &= \sum_{i=1}^{N_X} B^{\ell, m}_{X,i} (t) X^{\rm \ell, m}_i(q)  \, , ~~ X = \{A, \varphi\} .
\label{eq:surrogate_model}
\end{split}
\end{align}
Unlike Ref.~\cite{Field:2013cfa}, we construct a reduced basis representation for the 
waveform amplitudes and phases separately, instead of the waveforms themselves ~\cite{Purrer:2014}.
Here, the $\{B^{\ell m}_{X,i} \}_{i=1}^{N_X}$ are computed off-line from
the SpEC waveforms \cite{Field:2013cfa}. 
At a set of ${N_X}$ specially selected times 
$\{ T^{\ell m}_{X,i} \} _{i=1}^{N_X}$, which are the empirical interpolant nodes \cite{Maday:2009, Field:2013cfa},
the functions $X^{\ell m}_i(q) \approx X^{\ell m}(T^{\ell m}_{X,i}; q)$
approximate the parametric variation of the amplitudes and phases (via fitting).
A thorough discussion of surrogate model building steps
is presented in~\cite{Field:2013cfa}. 
When evaluating the surrogate at a particular mass ratio, the fits are  evaluated first
to determine the amplitudes and phases at their respective interpolating times $\{T^{\ell m}_{X,i} \}_{i=1}^{N_X}$. 
The remaining operations yield the surrogate model
prediction, $h_{\rm S}(t, \theta, \phi; q)$.

To find each $X^{\ell m}_i(q)$ 
we perform least-squares fits to the $22$ 
data points, $\{X^{\ell m}(T^{\ell m}_{X,i}; q_j) \}_{j=1}^{22}$.
All fits except odd $m$ mode amplitudes use $5$th degree polynomials in the symmetric mass ratio, $\nu = q/(1+q)^2$.
For odd $m$ modes, the 
amplitude approaches $0$ and its derivative with respect to $\nu$
diverges as $q\to 1$ (or $\nu\to 1/4$).
Consequently, we use
$A^{\ell m}_i(\nu) = \sum_{n=1/2, 1}^5 a^{\ell m}_n(1 - 4\nu)^n$
to account for this behavior.
The waveform phases of odd $m$ modes at $q=1$, which are undefined, are excluded when fitting for
each $\varphi^{\ell m}_i (q)$.

{\it Assessing surrogate errors--}
We next assess the surrogate's 
predictive quality. 
To quantify the error in the surrogate model itself, 
as opposed to its usage in a data analysis study,
we do {\it not} minimize
the errors over relative time and phase shifts here.

\begin{figure}
        \includegraphics[width=\columnwidth, trim=20 20 18 20]{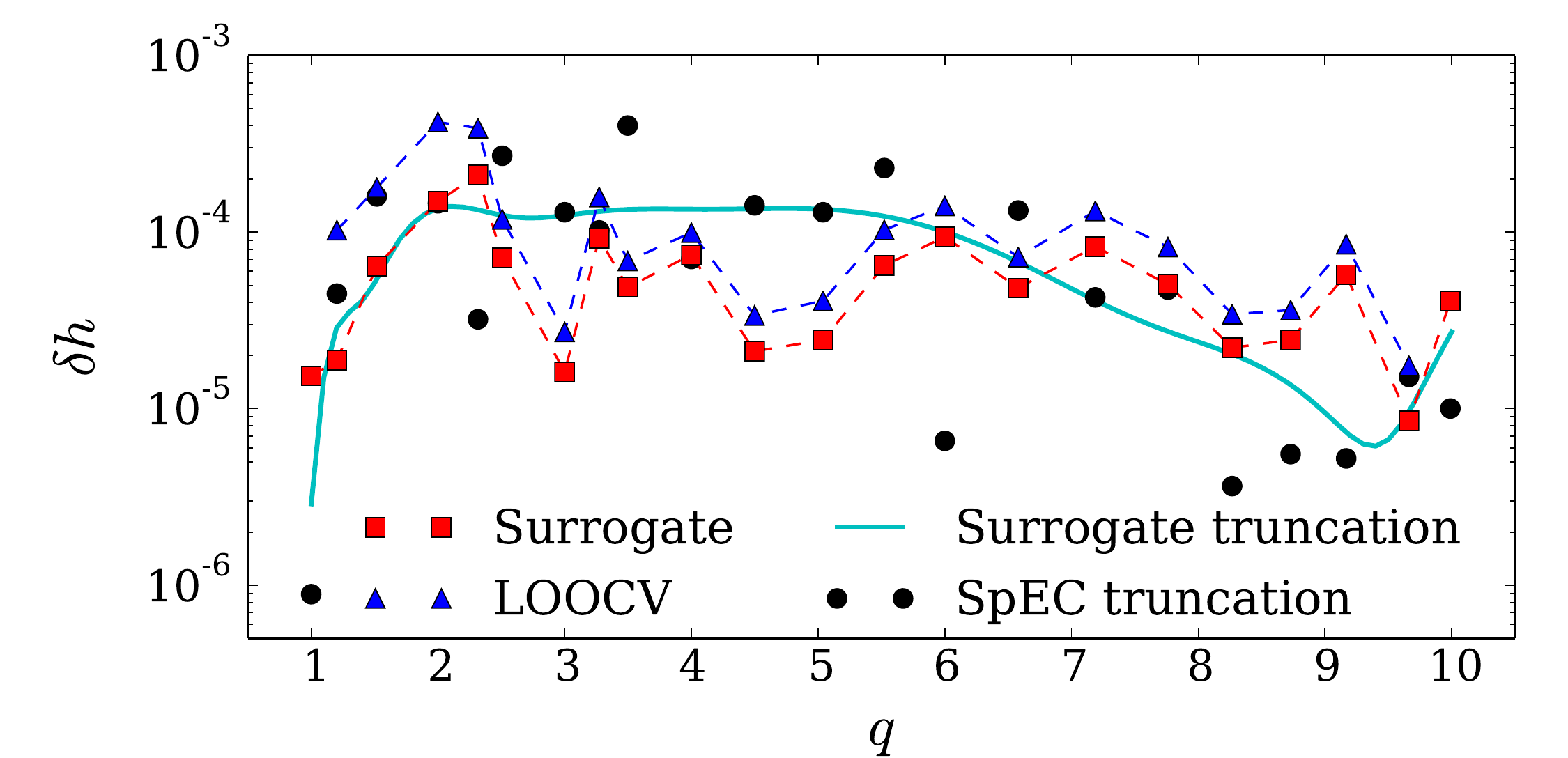}
        \caption{
        	Waveform differences between the two highest SpEC resolutions (black circles),
                surrogates built from the two highest SpEC resolutions (cyan line),
                the full surrogate and SpEC (red squares), and leave-one-out trial surrogates and SpEC (blue triangles).
                The largest surrogate error is for $q = 2$, for which the $(2,2)$ mode is shown in Fig.~\ref{fig:waveform}.
        }
        \label{fig:leaveOneOut}
\end{figure}

A first test is a consistency check to reproduce the $22$ input SpEC waveforms used to build the surrogate. 
These errors are shown 
in Fig.~\ref{fig:leaveOneOut}  (red squares)
and are comparable to or smaller than the \emph{largest} SpEC truncation errors (black circles).

A more stringent test is the 
leave-one-out cross-validation (LOOCV) study \cite{Hastie2001}. 
For each simulated mass ratio $q_i$, we build a temporary 
{\em trial surrogate} using the other $21$
waveforms, evaluate the trial surrogate at $q_i$, and compare
the prediction with the  
SpEC waveform for $q_i$. 
Hence, the trial surrogate's error at $q_i$ should 
serve as an
upper bound for the full surrogate 
trained on all $22$ waveforms.
Repeating this process for all possible $20$ LOOCV tests\footnote{We omit the smallest and largest mass ratios here 
as the corresponding trial surrogates would extrapolate to their values.}
results in Fig.~\ref{fig:leaveOneOut} (blue triangles).
Despite the $i$th trial surrogate having no information about 
the waveform at $q_i$, the errors remain comparable to
the \emph{largest} SpEC truncation errors. The LOOCV errors are typically
twice as large as the full surrogate ones 
confirming the former as bounds for the latter. Relative errors
for selected modes are shown in Table~\ref{tab:mode_err}. 
While weaker modes have larger relative errors, 
their power contribution is small enough that the
error in the full surrogate waveform, $\delta h$, is nearly identical
to the SpEC resolution error.

\begin{table}
\begin{tabular}{c | c c | c c || c | c c|c c}
\multirow{2}{*}{$(\ell,m)$} & \multicolumn{2}{c}{Surrogate} & \multicolumn{2}{|c||}{NR} & \multirow{2}{*}{$(\ell,m)$}& \multicolumn{2}{c}{Surrogate} & \multicolumn{2}{|c}{NR}\\
	& Max \!& \!\!Mean\! \!& Max \!& \!\!Mean &  & Max \!& \!\!Mean\! \!& Max \!& \!\!Mean\\
\hline
$(2,2)$ & 0.36 & 0.07 & 0.36 & 0.08 & $(3,2)$ & 100 & 17   & 1.7 & 0.43 \\
$(2,1)$ & 29   & 3.4  & 4.1  & 0.54 & $(4,4)$ & 7.4 & 2.2  & 20  & 2.1  \\
$(3,3)$ & 53   & 4.1  & 11   & 0.94 & All     & 0.42& 0.12 & 0.40 & 0.10  \\
\end{tabular}
\caption{
Relative mode errors, reported as $10^3 \times \| h_{\rm S}^{\ell,m}(q) - h^{\ell,m}(q) \|^2 / \| h^{\ell,m}(q) \|^2$,
from the leave-one-out surrogates.
Only those modes which contribute greater than 
$0.02$\% to the full waveform's time-domain power are used in the computation 
of the max and mean, except for `All' which is just $\delta h$. 
Our surrogate also includes the $(3, 1)$, $(4,[2,3])$, 
$(5,[3,4,5])$, $(6,[4,5,6])$, $(7,[5,6,7])$, and $(8,[7,8])$ modes.
Weaker modes typically have relative errors 
between $1\%$ and $35\%$.
\label{tab:mode_err}
}
\end{table}

A third test is to 
compare the surrogate waveforms to those of a second surrogate, built from the second highest resolution SpEC waveforms.
The resulting comparison 
is shown 
in Fig.~\ref{fig:leaveOneOut} (cyan line).
These errors are comparable to SpEC waveform truncation errors (black circles).
We find that the surrogate building process is robust to resolution differences. 
Furthermore, the surrogate can be improved using NR waveforms of higher accuracy.

We perform a final test and construct surrogates using 
the first $N$ selected mass ratios (from Table~\ref{tab:runs}) as input waveforms,
leaving $22-N$ mass ratios with which to test.
We find the total surrogate error
decreases exponentially with $N$ and is comparable to the SpEC truncation error
after using $15$ waveforms. Some modes (e.g., $(2, 2)$) are fully resolved after as few as $7$ waveforms.

\begin{figure}
    \includegraphics[width=\columnwidth, trim=35 0 20 20]{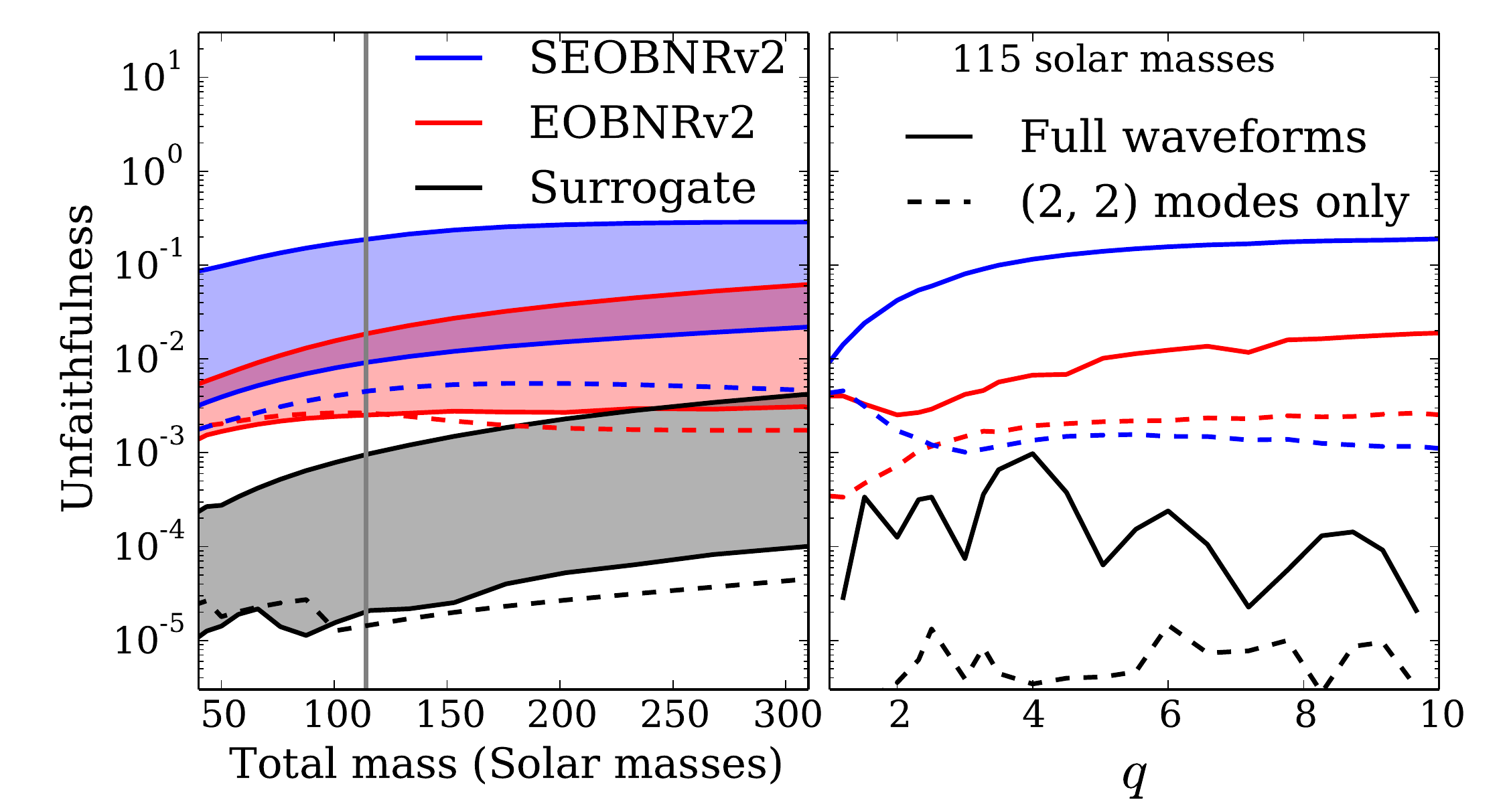}
    \caption{
Unfaithfulness, from Eq.~\eqref{eq:cheater}, comparing SpEC with our surrogate, EOBNRv2, and SEOBNRv2 models using all available $m \neq 0$ modes. 
Dashed lines show the unfaithfulness for $(2, 2)$ modes only.
All waveforms are Planck-tapered~\cite{McKechan:2010kp} for $t \in [-2750, -2500]M$ and $t \in [50, 90]M$.
For the full multi-modal waveforms, we maximize the unfaithfulness over $\theta$ and $\varphi$ for the worst-case scenario.
We use the ``$+$'' polarization, which is non-zero for all $(\theta,\varphi)$.
{\bf Left}: The shaded regions contain all $22$ mass ratios, while the dashed lines maximize over mass ratio.
The vertical grey line is the minimum total mass ($\approx$$115M_\odot$) ensuring all $(2, 2)$ modes start with
$\le 15$Hz at the end of the first tapering window. 
{\bf Right}: Unfaithfulness for a $115M_\odot$ binary.
	}
    \label{fig:EOBComparison}
\end{figure}

{\it Comparison to EOB--}
For data analysis purposes, we compare our surrogate with EOBNRv2 ~\cite{Pan:2011gk}
and SEOBNRv2 ~\cite{Taracchini:2012} models (generated from a current implementation\footnote{
We find that very small changes ($\sim$$10^{-12}$)
in the minimum frequency or the total mass can have unexpectedly large
changes 
in the unfaithfulness ($\sim$$10^{-4}$)}
in LAL~\cite{LAL}).
In Fig.~\ref{fig:EOBComparison}, we show the unfaithfulness
\begin{align}
1 - \max_{\delta \varphi, \delta t} \, {\rm Re} \!\! \int_{15{\rm Hz}}^{\infty} {\hskip-0.1in} df \, \frac{ \hat{\tilde{h}}_1(f; \theta, \varphi) \hat{\tilde{h}}_2^* (f; \theta, \varphi + \delta\varphi) e^{2\pi i f \delta t} }{ S_n(f) }  
\label{eq:cheater}
\end{align}
of the surrogate and the two EOB models against the NR waveforms. Here,
$\hat{\tilde{h}}$ is the
normalized Fourier transform of $h$ (such that a waveform's unfaithfulness with itself gives $0$), and $S_n(f)$ the advanced LIGO 
zero-detuned high power sensitivity noise curve~\cite{Shoemaker2009}.
The surrogate is more faithful than both EOB models for all cases considered.
Since SEOBNRv2 only provides $(2, \pm 2)$ modes, it performs
worst for large total masses
where additional modes become important.
All models predict the $(2, 2)$ mode with an unfaithfulness  
$< 1\%$ for $q \in [1, 10]$ at $115M_\odot$, however the EOB models are
limited by the availability of subdominant modes.

{\it Discussion--}
We have built a surrogate model for NR non-spinning BBH merger waveforms generated by SpEC.
On a standard 2015 single core computer, all $77$ modes with $2 \leq \ell \leq 8$ are evaluated in 
$\approx 0.5$ sec ($\approx 0.01$ sec for a single mode)
providing a factor of
$\sim 10^{6-8}$ speedup compared to SpEC. Importantly, this is achieved
with only a {\it small} loss in accuracy.
Like other data-driven modeling strategies, our surrogate is 
valid only 
within the training intervals, namely, $q \in [1,10]$ and $t/M \in [-2570, 100]$. 
Therefore, within the training intervals, our surrogate model generates BBH merger waveforms
that are equivalent to SpEC outputs up to numerical error and a small modeling error. 

NR surrogates can be used for multiple-query applications in gravitational wave data analysis such as 
detector-specific template-bank (re-)generation and parameter estimation. Our surrogate, and more generally the results of this paper, 
open up the exciting possibility of performing, for example, parameter estimation with multi-modal
NR waveforms (with hybridization, if needed).
Parameter estimation studies seeking to incorporate model error
may benefit from the surrogate's relatively
straightforward characterization and assessment of uncertainty from a combination of the surrogate's and  
SpEC's systematic and numerical errors.
We anticipate NR surrogate modeling to complement 
traditional strategies \cite{Buonanno99, Ajith-Babak-Chen-etal:2007, Damour2009a, Buonanno:2009qa, Pan:2009wj,
Santamaria:2010yb, Sturani:2010yv, Pan:2011gk, Abadie:2011kd, Damour:2012ky, Taracchini:2012} by 
providing unlimited high-fidelity approximations of NR waveforms with which to calibrate, refine and make comparisons.
Building NR surrogates of precessing BBH merger waveforms,
which may be modeled from the parameters specially selected in \cite{Blackman:2014maa},
offer a promising avenue for modeling the full $7$ dimensional BBH parameter space. 
The surrogate model described in this paper is available for download at \cite{SXSWebsiteSurrogate,gwsurrogate}.

We thank Mike Boyle, Alessandra Buonanno, Collin Capano, Jan Hesthaven, Jason Kaye, Geoffrey Lovelace, Lee Lindblom, 
Tom Loredo, Christian Ott, Yi Pan, Harald Pfeiffer, Rory Smith, and 
Nicholas Taylor for many useful discussions throughout this project.
This work was supported in part by 
NSF grants CAREER PHY-0956189, PHY-1068881, PHY-1005655,
PHY-1440083, PHY-1404569, and AST-1333520 to Caltech,
NSF grants PHY-1306125 and AST-1333129 to Cornell University, NSF grant PHY-1500818 to the University of California at San Diego, 
NSF grants PHY-1208861 and PHY-1316424 to the University of Maryland (UMD), 
NSERC of Canada, and the Sherman Fairchild Foundation.
Computations were performed on the Zwicky cluster at Caltech, which is supported by the Sherman Fairchild Foundation
and by NSF award PHY-0960291. Portions of this research were carried out at 
the Center for Scientific Computation and Mathematical Modeling cluster at UMD.

%

\end{document}